\documentclass[prd,showpacs,superscriptaddress,twocolumn,
floatfix,preprintnumbers,altaffilletter]{revtex4}
\usepackage{graphicx,url,amssymb,amsmath,longtable,rotating,color,units, 
wasysym,subfigure,epsfig,multirow}

\newcommand{\stochtrack}{{\em stochtrack}}
\newcommand{\stochsky}{{\em all-sky stochtrack}}
\newcommand{\burstegard}{{\em burstegard}}
\newcommand{\burstegardtab}{targeted seed-based}
\newcommand{\stochtracktab}{targeted seedless}
\newcommand{\stochskytab}{all-sky seedless}

\begin{document}

\title{Seedless clustering in all-sky searches for gravitational-wave transients}


\author{Eric~Thrane}
\email{ethrane@ligo.caltech.edu}
\affiliation{LIGO Laboratory, California Institute of Technology, MS 100-36,
Pasadena, CA, 91125, USA}

\author{Michael Coughlin}
\affiliation{Department of Physics, Harvard University, Cambridge, MA 02138, USA}

\date{\today}

\begin{abstract}
  The problem of searching for unmodeled gravitational-wave bursts can be thought of as a pattern recognition problem: how to find statistically significant clusters in spectrograms of strain power when the precise signal morphology is unknown.
  In a previous publication, we showed how ``seedless clustering'' can be used to dramatically improve the sensitivity of searches for long-lived gravitational-wave transients.
  In order to manage the computational costs, this initial analysis focused on externally triggered searches where the source location and emission time are both known to some degree of precision.
  In this paper, we show how the principle of seedless clustering can be extended to facilitate computationally-feasible, all-sky searches where the direction and emission time of the source are entirely unknown.
  We further demonstrate that it is possible to achieve a considerable reduction in computation time by using graphical processor units (GPUs), thereby facilitating more sensitive searches.
\end{abstract}

\maketitle

\section{Introduction}\label{intro}
Long-lived gravitational-wave transients (lasting $\gtrsim\unit[10]{s}$) constitute an interesting class of signals for second-generation detectors such as Advanced LIGO~\cite{aligo} and Advanced Virgo~\cite{avirgo}.
After reaching design sensitivity, Advanced LIGO expects to observe $\approx40$ binary neutron stars mergers and $\approx10$ neutron-star black-hole coalescences per year of science data~\cite{rates}.
The standard searches for compact binary coalescences rely on matched filter template banks; see, e.g.,~\cite{s6grb,s6lowmass}.
More exotic sources of long-lived transients, including emission from rotational instabilities in protoneutron stars~\cite{pirothrane12,piro:11,piro:07,corsi} and black-hole accretion disk instabilities~\cite{kiuchi,vanputten:01,vanputten:08}, cannot be accurately modeled owing to theoretical uncertainties.
However, searches for long-lived bursts~\cite{lgrb,stamp,stochtrack} can be employed when a matched filter search is not possible.
(There is a rich literature on short, sub-second gravitational-wave bursts and the different detection strategies available to detect them, but we focus here on long-lived transients.)

In a cross-correlation search such as~\cite{lgrb,stamp}, the detection of gravitational waves can be thought of as a pattern recognition problem.
The goal is to find tracks of excess strain cross-power, which appear as brighter-than-expected pixels on a signal-to-noise ratio spectrogram ($ft$-map).
In a previous work~\cite{stochtrack}, we described how ``seedless clustering'' can be used to significantly enhance the sensitivity of searches for long-lived transients when a trusted matched filter template bank is not available.
We review the details of seedless clustering in Sec.~\ref{algorithm}, but the basic idea is to integrate along many different cleverly chosen paths in a signal-to-noise-ratio spectrogram.
This is in contrast to seed-based clustering algorithms which form clusters from bright spectrogram pixels called ``seeds.''

The advantage of seedless clustering is most pronounced for long and weak signals~\cite{stochtrack}.
For the waveforms considered in~\cite{stochtrack}, we found that seedless clustering can detect a gravitational-wave signal (at a fixed false-alarm and false dismissal rate) at a distance between $1.5$--$2\times$ further than a seed-based clustering algorithm.
This corresponded to an increased detection volume of $4.2$--$7.4\times$.

One of the challenges associated with seedless clustering is that it is, as a rule of thumb, more computationally expensive than seed-based alternatives.
In~\cite{stochtrack}, we focused on applications to {\em targeted} searches, in which the sky location is tightly constrained and the time of the event is known to exist in some ``on-source'' window, thereby saving the extra computational cost associated with searching many sky positions and emission times.

In this work, we show how the seedless clustering formalism from~\cite{stochtrack} can be extended to a high-sensitivity, computationally-efficient, {\em all-sky} search for long-lived gravitational waves from arbitrary sky locations.
There are two innovations which make this possible.
First, by introducing a new random phase factor, we show that it is possible to efficiently scan the entire sky with a seedless clustering algorithm.
Second, we take advantage of recent advances in computing to carry out our computations on graphical processor units (GPUs).
Seedless clustering algorithms are ``embarrassingly parallel''~\cite{parallel}, which allows them to exploit the highly parallel architecture of GPUs.
We show that an all-sky search with seedless clustering is both computationally feasible and more sensitive than a seed-based algorithm.
We outline the computational requirements for a realistic search and demonstrate the advantage of carrying out computations on GPUs.

The remainder of the paper is organized as follows.
In Sec.~\ref{challenges}, we describe some of the general features and challenges of an all-sky transient search.
In Sec.~\ref{algorithm}, we describe \stochsky, an all-sky algorithm which employs seedless clustering.
In Sec.~\ref{results}, we present the results of a sensitivity study comparing \stochsky\ to a seed-based algorithm.
In Sec.~\ref{computing}, we describe the computational resources required for realistic searches and compare the algorithms' performance on CPUs and GPUs.
In Sec.~\ref{conclusions}, we offer concluding remarks and suggest directions for future research.

\section{The challenges of all-sky radiometry}\label{challenges}
In this section, we outline some of the general features of an all-sky transient search built on the principle of radiometry---using the time delay between two detectors to search data associated with a specific direction in the sky.
We begin with strain time series $s_I(t')$ and $s_J(t')$ from detectors $I$ and $J$, which are separated by a displacement $\Delta\vec{x}$.
The data are split into segments (typically with a duration of $\approx\unit[1]{s}$) and Fourier-transformed to create complex-valued strain spectrograms: $\tilde{s}_I(t;f)$ and $\tilde{s}_J(t;f)$.
Note that $t'$ refers to sampling time whereas $t$ refers to segment start time.

Following~\cite{stamp,stochtrack,lgrb}, the signal-to-noise ratio spectrogram can be written as:
\begin{equation}\label{eq:rho}
  \rho(t;f|\hat\Omega) = \text{Re}\left[
    \lambda(t;f)
    e^{2\pi i f \Delta\vec{x}\cdot\hat\Omega/c}
    \tilde{s}_I^*(t;f) \tilde{s}_J(t;f)
    \right] .
\end{equation}
Here, $e^{2\pi i f \Delta\vec{x}\cdot\hat\Omega/c}$ is a phase factor, which takes into account the time delay between detectors $I$ and $J$; $c$ is the speed of light.
The phase factor rotates the cross-power signal in the complex plane so as to be real and positive.
The $\lambda(t;f)$ term is a normalization factor, which uses neighboring segments to estimate the background~\footnote{
The $\lambda(t;f)$ factor may also include a direction-dependent phase factor taking into account the relationship between the $+$ and $\times$ polarizations of an elliptically polarized source.
For the sake of simplicity, we do not include this additional phase factor.
We expect the sensitivity to improve marginally by adding this phase factor by incorporating additional polarization information, though, at an increased computational cost.
}.
Precise definitions of $\rho(t;f|\hat\Omega)$ and $\lambda(t;f)$ are provided in Appendix~\ref{formalism}.
If the source direction $\hat\Omega$ is known, for example, from an electromagnetic trigger (see~\cite{stochtrack}), then it is straightforward to apply to the appropriate phase factor.
When no electromagnetic trigger is available, it is necessary to search over multiple directions.

Consider the case where the source is located at $\hat\Omega$ but the filter is chosen for the direction $\hat\Omega'$, which introduces a timing error of:
\begin{equation}
  \Delta\tau = \Delta\vec{x}\cdot(\hat\Omega-\hat\Omega')/c .
\end{equation}
On average, the timing error reduces the signal-to-noise ratio by
\begin{equation}\label{eq:ratio}
  R \equiv
  \left\langle\rho(t;f|\hat\Omega')\right\rangle
  /
  \left\langle\rho(t;f|\hat\Omega)\right\rangle 
  = \cos(2\pi f \Delta\tau) 
  \equiv \cos(\delta) .
\end{equation}
Inspecting Eq.~\ref{eq:ratio}, we can infer the qualitative features of a signal in a $\rho(t;f|\hat\Omega')$ spectrogram characterized by a timing error $\Delta\tau$.
For small values of $\delta$, the apparent signal will be weaker than it would in the absence of a timing error.
This is because some of the cross-power in Eq.~\ref{eq:rho} leaks into the imaginary direction.
As $\delta$ crosses $\pi/2$, the signal vanishes entirely before reappearing as negative signal-to-noise ratio.

\begin{figure*}[hbtp!]
  \begin{tabular}{cc}
    \psfig{file=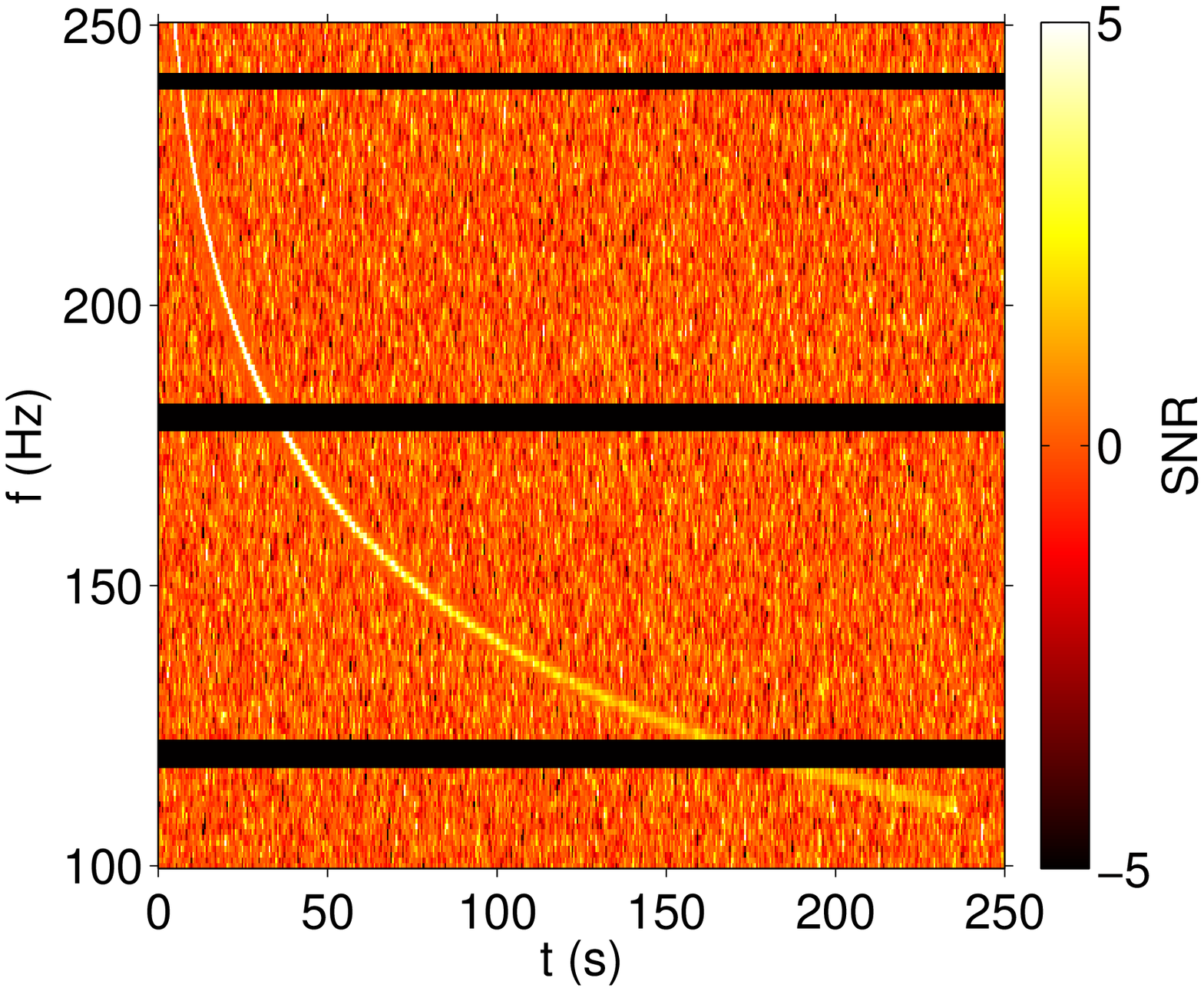, height=2.8in} & 
    \psfig{file=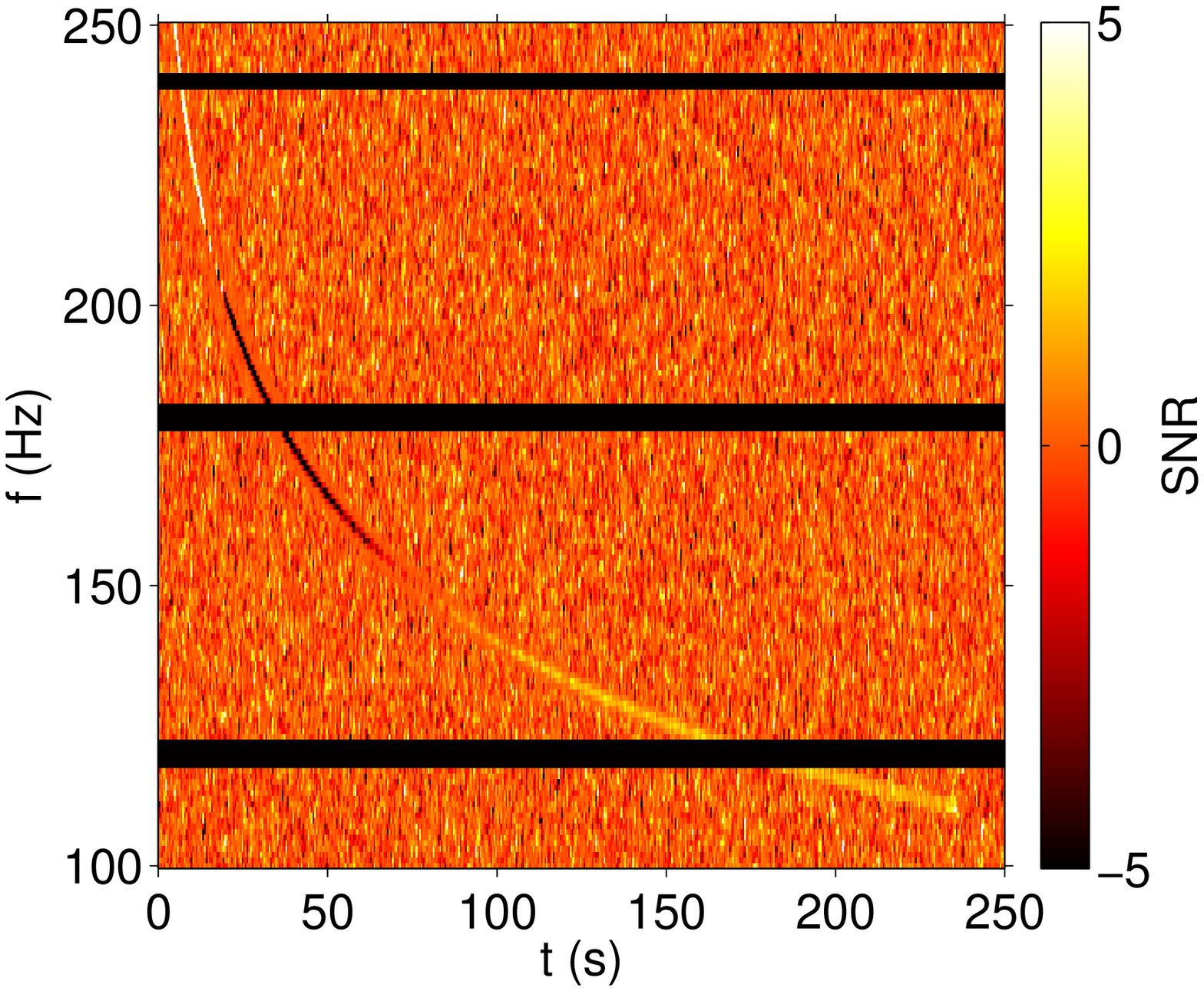, height=2.8in} \\
    \psfig{file=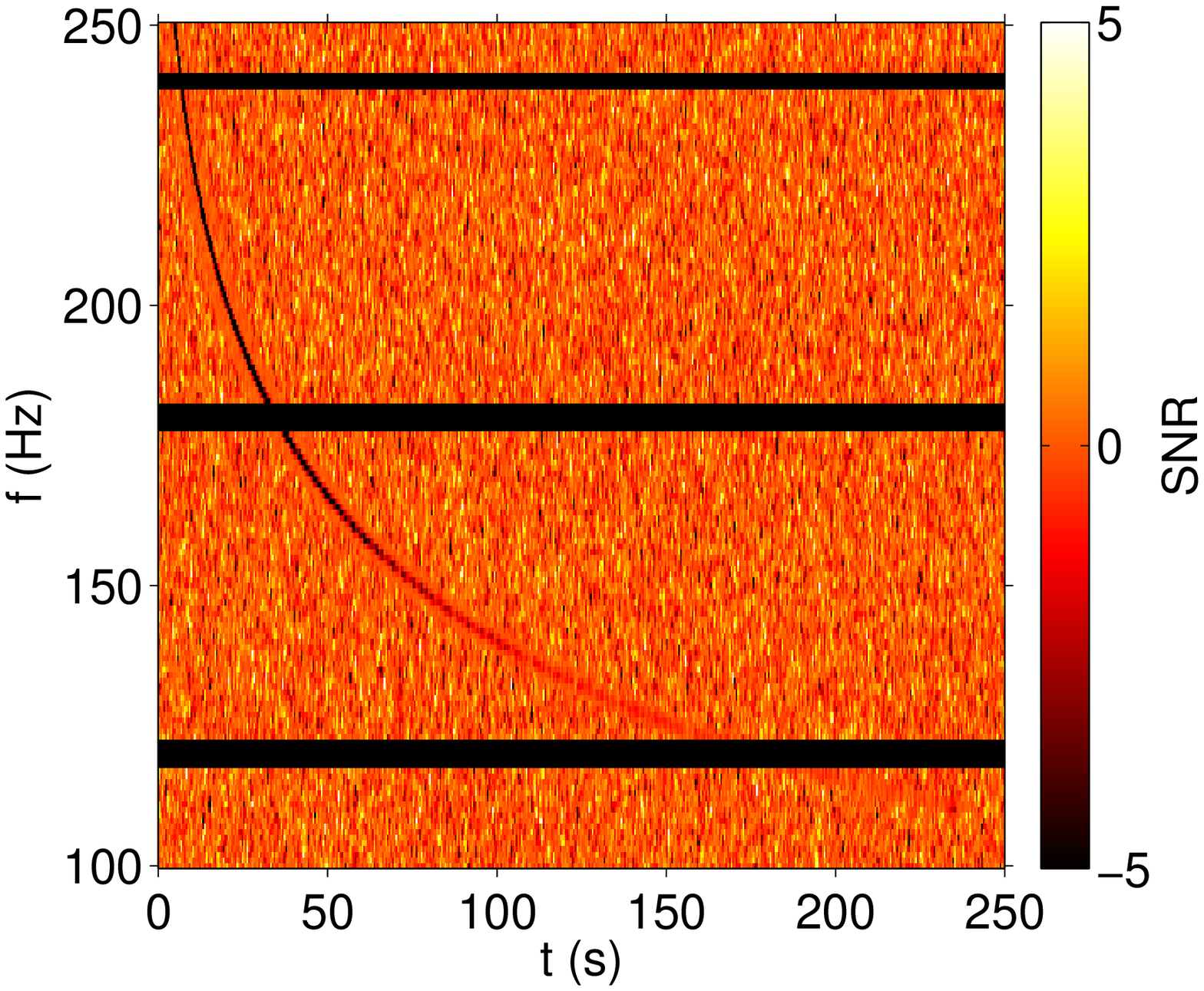, height=2.8in} &
    \psfig{file=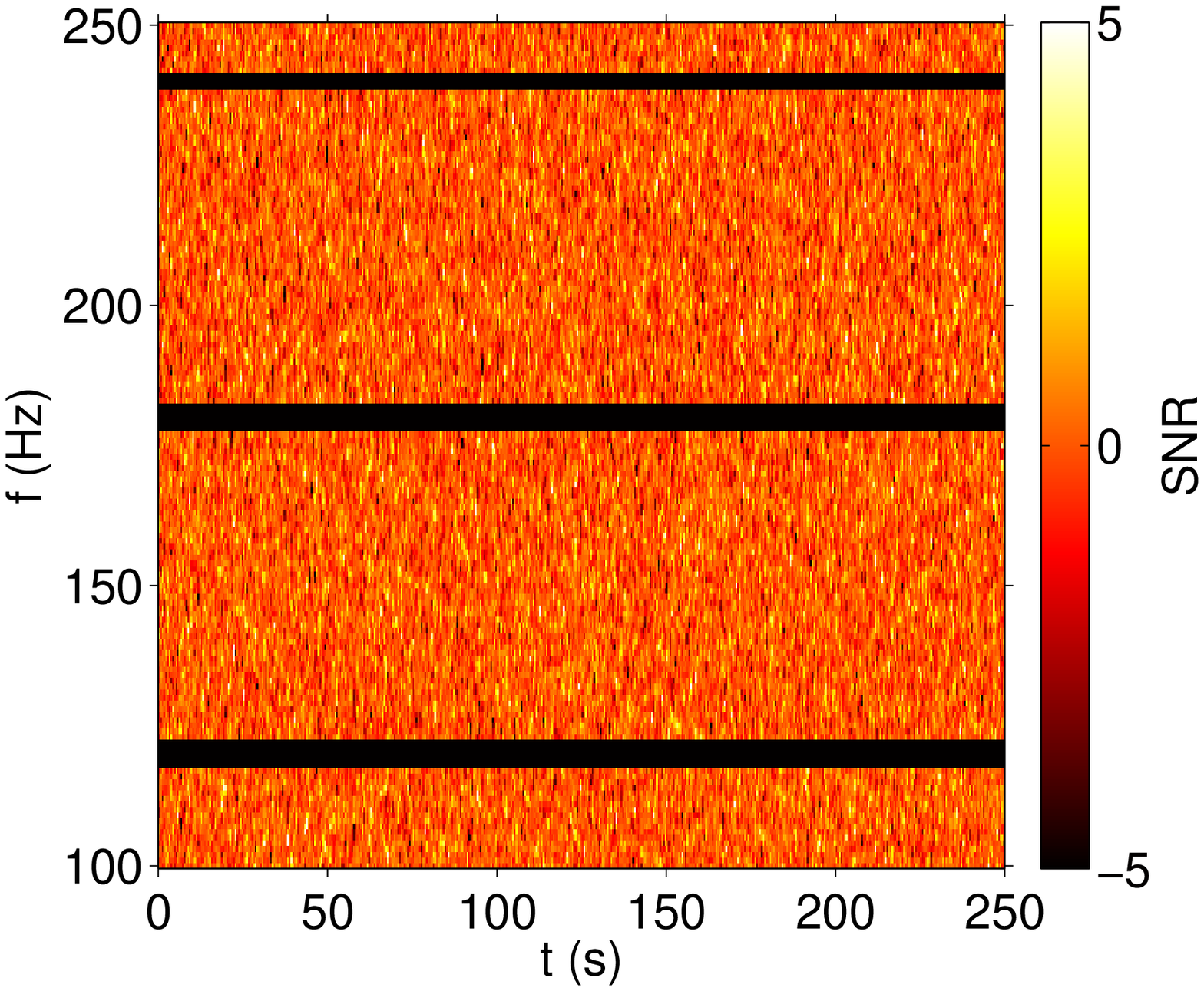, height=2.8in}
    \end{tabular}
  \caption{
    The effect of filter mismatch.
    Top-left: $\text{SNR}$ spectrogram showing an accretion disk instability signal (ADI~2) in Advanced LIGO Monte Carlo noise and obtained using the correct filter.
    The signal appears as a whitish-yellow track indicating positive $\text{SNR}$.
    The signal has been made very loud ($d=\unit[50]{Mpc}$) for illustrative purposes.
    The black horizontal lines are notches due to instrumental artifacts.
    Top-right: the same signal using an incorrect filter, i.e., the search direction does not match the source direction.
    The mismatch causes alternating stripes of positive (yellow) and negative (reddish black) $\text{SNR}$.
    At the turning points, where the $\text{SNR}$ switches from positive to negative, the filtered cross-power signal is imaginary, and so the $\text{SNR}$ (proportional to the real part of the filtered cross power) is approximately zero.
    Bottom-left: the same signal using a different incorrect filter.
    In this case, the signal appears as purely negative.
    Bottom-right: the same signal, using the correct filter, but much further away ($D=\unit[340]{Mpc}$).
    It is all but impossible to see the track with the naked eye, but it can nonetheless be detected by \stochsky\ with $\text{FAP}<0.1\%$ without knowledge of the true direction or the signal morphology.
  }
  \label{fig:stochsky}
\end{figure*}

Graphically, large timing errors produce characteristic stripes in $\rho(t;f|\hat\Omega)$ spectrograms; see Fig.~\ref{fig:stochsky}.
The bandwidth of each stripe is given by $1/4\Delta\tau$.
The minimum stripe size is $\Delta f_\text{min} = 1/4\Delta\tau_\text{max}$, where $\Delta\tau_\text{max}$ is the travel time between detectors $I$ and $J$.
For the two LIGO detectors, $\Delta f_\text{min} \approx \unit[25]{Hz}$.

We can define a tolerance for the maximum possible timing error by requiring that we observe no less than, say, $R=90\%$ of the signal-to-noise ratio.
It follows that
\begin{equation}
  \Delta\tau < \frac{1}{2 \pi f}\cos^{-1}(R) .
\end{equation}
As frequency increases, the tolerable timing error decreases.
For signals in the most sensitive part of LIGO's band $f\approx\unit[100]{Hz}$, the $R\geq90\%$ timing tolerance is $\Delta\tau\leq\unit[720]{\mu s}$.
For high-frequency signals near $f\approx\unit[1000]{Hz}$, it is $\Delta\tau\leq\unit[72]{\mu s}$.

To summarize, radiometry relies on the use of a phase factor to characterize the time delay between two detectors (a function of source sky position).
If the assumed sky position is incorrect, gravitational-wave cross-power leaks from the positive real direction into imaginary and/or negative components, which, in turn, leads to a reduced signal-to-noise ratio.
It may therefore be necessary to search many directions (with many time delays) in order to observe the signal with an acceptable signal-to-noise ratio.

\section{Algorithm}\label{algorithm}
In this section, we describe the details of an algorithm which uses seedless clustering to search for transient signals from all directions on the sky.
We call it \stochsky.
We begin with a brief review of the \stochtrack\ algorithm~\cite{stochtrack}, which will serve as a foundation on which to build.

The goal is to find the most significant cluster $\Gamma$ as determined by the value of the detection statistic~\footnote{
We note that alternative definitions of $\text{SNR}_\text{tot}$ (with a different weighting scheme) are also possible; see~\cite{stochtrack}.
},
\begin{equation}\label{eq:sum}
  \text{SNR}_\text{tot} \equiv
  \frac{1}{N}
  \sum_{\left\{t;f\right\}\in\Gamma} \rho(t;f|\hat\Omega)
\end{equation}
where $\rho(t;f|\hat\Omega)$ is defined in Eq.~\ref{eq:rho} and $N$ is the number of pixels in $\Gamma$.

In any seedless clustering algorithm, $\Gamma$ is determined a priori by some set of rules (as opposed to by the data itself).
In the \stochtrack\ algorithm~\cite{stochtrack}, $\Gamma$ is chosen randomly from the set of quadratic B\'ezier curves~\cite{bezier} subject to the constraint that the curve persists for a duration $t_\text{min}$.
(Other parameterizable curves such as spline can be used as well.)
Each randomly selected B\'ezier curve is referred to as a ``template.''
Each B\'ezier template is described by three time-frequency control points: $P_0$ $(t_\text{start}, f_\text{start})$, $P_1$ $(t_\text{mid}, f_\text{mid})$, and $P_2$ $(t_\text{end}, f_\text{end})$.
The control points form a curve parameterized by $\xi=[0,1]$:
\begin{equation}
  \left(\begin{array}{c} t(\xi) \\ f(\xi) \end{array}\right) =
  (1-\xi)^2 P_0 + 2(1-\xi)\xi P_1 +
  \xi^2 P_2 .
\end{equation}

In order for the algorithm to have a high probability of guessing a close approximation to the true signal, many templates must be used.
Fortunately, each template can be quickly generated from just six random numbers.
By working with {\em arrays} of B\'ezier curves, \stochtrack\ is able to carry out the sum in Eq.~\ref{eq:sum} for a large number of templates in parallel.
As we shall see below, the parallel nature of the calculation lends itself to the use of GPUs.

The number of templates is denoted $T$.
For practical applications, it is typically chosen to be $T={\cal O}(10^5--10^8)$.
In~\cite{stochtrack}, we described a default search with $T=2\times10^6$ and a deep search (denoted \stochtrack\ $10\times$) with $T=2\times10^7$~\footnote{
In~\cite{stochtrack}, the default and deep search are said to use $T=2\times10^7$ and $T=2\times10^8$ respectively.
We believe the correct numbers are actually $T=2\times10^6$ and $T=2\times10^7$ as stated here.
}.

The \stochsky\ algorithm builds on the foundation of \stochtrack.
First, we introduce ``complex signal-to-noise ratio'':
\begin{equation}\label{eq:german_p}
  \mathfrak{p}(t;f) = 
    \lambda(t;f)
    \tilde{s}_I^*(t;f) \tilde{s}_J(t;f) .
\end{equation}
This is necessary in order to preserve the complex phase information that encodes the direction of the source.
Note that, unlike $\rho(t;f|\hat\Omega)$, $\mathfrak{p}(t;f)$ is not defined for a particular direction.

Next, in addition to the six random control points, we add an additional random variable $\Delta\tau$ corresponding to the time delay between the detectors, which, as we saw in Sec.~\ref{challenges}, is a proxy for sky location.
If we assume that the sky location of each transient is drawn from an approximately isotropic distribution, then the probability density function for time delay is a simple uniform distribution between $\pm\Delta\tau_\text{max}$.

Finally, we modify Eq.~\ref{eq:sum} to be 
\begin{equation}\label{eq:german_sum}
  \text{SNR}_\text{tot} \equiv
  \frac{1}{N}
  \sum_{\left\{t;f\right\}\in\Gamma} 
  \text{Re}\left[
    e^{2\pi i f \Delta\tau} \mathfrak{p}(t;f) 
    \right] .
\end{equation}
The new sum described in Eq.~\ref{eq:german_sum} is carried out for many randomly selected clusters $\Gamma$, each with a randomly selected time delay $\Delta\tau$.
By including a random time delay, the algorithm tries to guess not only the spectrographic shape of the signal, but also the appropriate phase factor that will minimize the timing error stripes shown in Fig.~\ref{fig:stochsky}.
Minimizing the timing error maximizes $\text{SNR}_\text{tot}$.

The addition of a new random variable comes at a cost.
First, the \stochsky\ algorithm will converge less quickly than \stochtrack\ due to its expanded parameter space.
Second, even if we imagine setting $T\rightarrow\infty$, \stochsky\ templates span a larger space than the templates used in \stochtrack, and so \stochsky\ must contend with comparatively higher background.
That said, we find that the extra cost is small.
In the next section, we show that, for several signal models, \stochsky\ achieves a sensitivity which is only slightly less than \stochtrack, while searching a significantly expanded signal space.

\section{Sensitivity Study}\label{results}
In this section, we describe a study to determine the sensitivity of \stochsky\ to four different long-lived test waveforms using Monte Carlo and recolored LIGO noise.
In order to compare with the baseline sensitivity of \stochtrack, we use the same four waveforms used in~\cite{stochtrack}: two fallback accretion signals~\cite{pirothrane12} abbreviated FA~1 and FA~2, and two accretion-disk instability waveforms~\cite{lucia}, abbreviated ADI~1 and ADI~2.
The waveforms span durations of $25$--$\unit[230]{s}$ and range in frequency from $110$--$\unit[1530]{Hz}$.
Additional details about the waveforms and the parameters used to generate them are provided in Appendix~\ref{parameters}.
Additional information about the models behind the waveforms is available in~\cite{pirothrane12} and~\cite{lucia}; see also~\cite{piro:11,piro:07,corsi,kiuchi,vanputten:01,vanputten:08}.

Each waveform is injected into either Monte Carlo or recolored noise (initial LIGO noise which has been recolored to match the design sensitivity of Advanced LIGO while preserving non-stationary noise artifacts)~\footnote{
The data are taken in between GPS times 822917487 and 847549782.
}.
The data are processed to form a complex signal-to-noise ratio spectrogram $\mathfrak{p}(t;f)$ (see Eq.~\ref{eq:german_p}).
Following~\cite{stochtrack}, the ADI waveforms are analyzed in a band between $100$--$\unit[250]{Hz}$ while the FA waveforms are analyzed in a band between $700$--$\unit[1600]{Hz}$.
The spectrogram resolution is $\unit[1]{s}\times\unit[1]{Hz}$ except for FA~1, for which we use $\unit[0.5]{s}\times\unit[2]{Hz}$.
Each spectrogram corresponds to $\unit[250]{s}$ of data.
Data segments are constructed with $50\%$-overlapping Hann windows.

We characterize the sensitivity in terms of a detection distance, defined as the distance to which \stochsky\ can detect a source with a false-alarm probability $\text{FAP}<0.1\%$ and a false-dismissal probability $\text{FDP}=50\%$.
We perform two series of tests.
First, in order to compare \stochsky\ with \stochtrack, we inject each signal with an optimal orientation (face-on) and in an optimal sky location (where the detectors are most sensitive).
The true source location is provided as input to \stochtrack\ (and to the seed-based clustering algorithm, \burstegard~\cite{burstegard}), but the \stochsky\ algorithm is not provided any information about the true location of the source.

We show \stochtrack\ results (from~\cite{stochtrack}) for the default search ($T=2\times10^6$) and for the deep search ($T=2\times10^7$), which is labeled: \stochtrack\ $10\times$.
We compare these to new results obtained with the default \stochsky\ ($T=2\times10^6$) and \stochsky\ $10\times$ ($T=2\times10^7$).

In the second series of tests, we inject signals at random sky locations $(\text{ra},\text{dec})$ (chosen from an isotropic distribution) and with random inclination and polarization angles $(\iota,\psi)$.
We expect that the detection distance for signals recovered with random values of $(\text{ra},\text{dec},\iota,\psi)$ will be $\approx60\%$ of what is achieved for optimal sources based on the antenna response of our two-detector network.

Our hypothetical network consists of the Advanced LIGO detectors in Hanford, WA (H1) and Livingston, LA (L1)~\cite{aligo}.
We assume both detectors are operating at design sensitivity.

The results of the study are summarized in Tables~\ref{tab:mc} and~\ref{tab:rn} for Monte Carlo and recolored noise respectively.
The Monte Carlo and recolored noise are processed identically except we apply a glitch identification~\cite{stamp_glitch} cut when analyzing recolored noise~\footnote{In order to apply the algorithm from~\cite{stamp_glitch}, we assume that the source is optimally oriented with an optimal sky position.}.
For optimally oriented sources injected into Monte Carlo noise, we find that the \stochsky\ $10\times$ can see sources $120$--$180\%$ further than the seed-based \burstegard, even though the \burstegard\ algorithm is given the known sky location whereas \stochsky\ is not.
This corresponds to increased detection volume of $180$--$560\%$.
For recolored noise, the improvement is $100$--$180\%$ in distance and $100$--$560\%$ in volume.

Repeating the Monte Carlo analysis with the computationally cheaper default version of \stochsky\ ($T=2\times10^7$), we obtain distances of $110$--$160\%$ times the distances obtained using \burstegard.
For recolored noise, these distances are $75$--$160\%$ times the values obtained using \burstegard.
Note that while \burstegard\ can detect the FA~1 waveform in recolored noise at greater distances than the default version of \stochsky, this is very likely because the \burstegard\ algorithm is supplied with the true source location.
In an apples-to-apples comparison, seedless clustering using the default \stochtrack\ is more sensitive than \burstegard~\cite{stochtrack}.

Thus, the fact that \burstegard\ can detect FA~1 signals in recolored noise at greater distances than \stochsky\ is telling us that it is very useful to know where in the sky to look when trying to find FA~1 waveforms in recolored noise.
This is, perhaps, not surprising since the FA~1 waveform is shorter and spans a greater bandwidth than the other waveforms we consider.
Shorter signals are more prone to resemble non-stationary noise.
Signals with larger bandwidth are more prone to loss of signal from phase factor mismatch (see Eq.~\ref{eq:german_sum}).

We also present detection distance for sources with random values of $(\text{ra},\text{dec},\iota,\psi)$, which are between $50$--$68\%$ of the values obtained for the case of an optimal source.

\begin{table}
  \begin{tabular}{|c|c|c|c|c|c|}
    \hline
    \multicolumn{1}{|c|}{waveform} &
    \multicolumn{1}{|c|}{algorithm} &
    \multicolumn{2}{c|}{distance} &
    \multicolumn{1}{c|}{volume} \\
    & & absolute & \% & \% \\\hline
    \multirow{4}{*}{ADI 1}
    & \burstegardtab & $\unit[370]{Mpc}$ & $100$ & $100$ \\
    & \stochtracktab & $\unit[540]{Mpc}$ & $150$ & $320$ \\
    & \stochtracktab\ $10\times$ & $\unit[590]{Mpc}$ & $160$ & $420$ \\
    & \stochskytab & $\unit[490]{Mpc}$ & $130$ & $240$ \\
    & \stochskytab\ $10\times$ & $\unit[540]{Mpc}$ & $150$ & $320$ \\
    & ...w/ random $(\text{ra},\text{dec},\iota,\psi)$ & $\unit[290]{Mpc}$ & --- & --- \\\hline
    \multirow{4}{*}{ADI 2}
    & \burstegardtab & $\unit[190]{Mpc}$ & $100$ & $100$ \\
    & \stochtracktab & $\unit[340]{Mpc}$ & $180$ & $560$ \\
    & \stochtracktab\ $10\times$ & $\unit[370]{Mpc}$ & $200$ & $740$ \\
    & \stochskytab & $\unit[310]{Mpc}$ & $160$ & $430$ \\
    & \stochskytab\ $10\times$ & $\unit[340]{Mpc}$ & $180$ & $560$ \\
    & ...w/ random $(\text{ra},\text{dec},\iota,\psi)$ & $\unit[200]{Mpc}$ & --- & --- \\\hline
    \multirow{3}{*}{FA 1}
    & \burstegardtab & $\unit[17]{Mpc}$ & $100$ & $100$ \\
    & \stochtracktab &  $\unit[29]{Mpc}$ & $150$ & $320$ \\
    & \stochtracktab\ $10\times$ & $\unit[35]{Mpc}$ & $180$ & $560$ \\
    & \stochskytab & $\unit[22]{Mpc}$ & $110$ & $130$ \\
    & \stochskytab\ $10\times$ & $\unit[24]{Mpc}$ & $120$ & $180$ \\
    & ...w/ random $(\text{ra},\text{dec},\iota,\psi)$ & $\unit[12]{Mpc}$ & --- & --- \\\hline
    \multirow{3}{*}{FA 2}
    & \burstegardtab & $\unit[25]{Mpc}$ & $100$ & $100$ \\
    & \stochtracktab & $\unit[36]{Mpc}$ & $150$ & $320$ \\
    & \stochtracktab\ $10\times$ & $\unit[40]{Mpc}$ & $160$ & $420$ \\
    & \stochskytab & $\unit[30]{Mpc}$ & $120$ & $180$ \\
    & \stochskytab\ $10\times$ & $\unit[36]{Mpc}$ & $150$ & $320$ \\
    & ...w/ random $(\text{ra},\text{dec},\iota,\psi)$ & $\unit[22]{Mpc}$ & --- & --- \\\hline
  \end{tabular}
  \caption{
    Comparing \stochsky\ (all-sky seedless) sensitivity to \stochtrack\ (tareted seedless) and \burstegard\ (targeted seed-based) results from~\cite{stochtrack} using Monte Carlo noise.
    Both \burstegard\ and \stochtrack\ are provided the true sky location as an input, while \stochsky\ searches over the entire sky.
    By default, \stochtrack\ and \stochsky\ perform $T=2\times10^6$ templates.
    The deep-search versions \stochtrack\ and \stochsky, denoted $10\times$, use $T=2\times10^7$ templates.
    ``Distance'' refers to the distance at which a source is detected with false alarm probability $=0.1\%$ and false dismissal probability $=50\%$.
    We list both the absolute distance in $\unit[]{Mpc}$ and the \% relative to the targeted seed-based algorithm.
    The ADI waveforms have been scaled assuming an energy budget of $E_\text{GW}=0.1 M_\odot$.
    Volume is given in \% relative to the targeted seed-based algorithm.
    All the results are for optimally oriented sources in an optimal sky location except for entries marked ``...w/ random $(\text{ra},\text{dec},\iota,\psi)$,'' which are an average over random sky locations and orientations.
    \label{tab:mc}
  }
\end{table}

\begin{table}
  \begin{tabular}{|c|c|c|c|c|c|}
    \hline
    \multicolumn{1}{|c|}{waveform} &
    \multicolumn{1}{|c|}{algorithm} &
    \multicolumn{2}{c|}{distance} &
    \multicolumn{1}{c|}{volume} \\
    & & absolute & \% & \% \\\hline
    \multirow{4}{*}{ADI 1}
    & \burstegardtab & $\unit[330]{Mpc}$ & $100$ & $100$ \\
    & \stochtracktab & $\unit[540]{Mpc}$ & $160$ & $420$ \\
    & \stochtracktab\ $10\times$ & $\unit[540]{Mpc}$ & $160$ & $420$ \\
    & \stochskytab & $\unit[450]{Mpc}$ & $130$ & $240$ \\
    & \stochskytab\ $10\times$ & $\unit[450]{Mpc}$ & $130$ & $240$ \\
    & ...w/ random $(\text{ra},\text{dec},\iota,\psi)$ & $\unit[280]{Mpc}$  & --- & --- \\\hline
    \multirow{4}{*}{ADI 2}
    & \burstegardtab & $\unit[170]{Mpc}$ & $100$ & $100$ \\
    & \stochtracktab & $\unit[310]{Mpc}$ & $180$ & $560$ \\
    & \stochtracktab\ $10\times$ & $\unit[340]{Mpc}$ & $200$ & $740$ \\
    & \stochskytab\ & $\unit[280]{Mpc}$ & $160$ & $420$ \\
    & \stochskytab\ $10\times$ & $\unit[310]{Mpc}$ & $180$ & $560$ \\
    & ...w/ random $(\text{ra},\text{dec},\iota,\psi)$ & $\unit[210]{Mpc}$  & --- & --- \\\hline
    \multirow{3}{*}{FA 1}
    & \burstegardtab & $\unit[22]{Mpc}$ & $100$ & $100$ \\
    & \stochtracktab &  $\unit[32]{Mpc}$ & $150$ & $320$ \\
    & \stochtracktab\ $10\times$ & $\unit[35]{Mpc}$ & $160$ & $420$ \\
    & \stochskytab & $\unit[16]{Mpc}$ & $75$ & $42$ \\
    & \stochskytab\ $10\times$ & $\unit[22]{Mpc}$ & $100$ & $100$ \\
    & ...w/ random $(\text{ra},\text{dec},\iota,\psi)$ & $\unit[11]{Mpc}$  & --- & --- \\\hline
    \multirow{3}{*}{FA 2}
    & \burstegardtab & $\unit[25]{Mpc}$ & $100$ & $100$ \\
    & \stochtracktab & $\unit[40]{Mpc}$ & $160$ & $420$ \\
    & \stochtracktab\ $10\times$ & $\unit[44]{Mpc}$ & $180$ & $560$ \\
    & \stochskytab\ & $\unit[30]{Mpc}$ & $120$ & $180$ \\
    & \stochskytab\ $10\times$ & $\unit[33]{Mpc}$ & $130$ & $230$ \\
    & ...w/ random $(\text{ra},\text{dec},\iota,\psi)$ & $\unit[21]{Mpc}$  & --- & --- \\\hline
  \end{tabular}
  \caption{
    The same as Table~\ref{tab:mc} except we utilize recolored noise from initial LIGO.
    An unphysical time shift is applied to spoil the coherence of any actual gravitational-wave signals that might have been present.
  }
  \label{tab:rn}
\end{table}

\section{Computing}\label{computing}
The results from Section~\ref{results} were obtained using graphical processor units (GPUs) on the LIGO Data Grid.
In this section, we document how GPUs provide an efficient architecture for carrying out \stochtrack\ and \stochsky\ calculations.
We compare the performance of the algorithm using both GPUs and CPUs.
We utilize Kepler GK104s GPUs, which are capable of peak single precision floating point performance of $\unit[4.6]{Tflops}$ according to the manufacturer.
Each GPU card has $\unit[4]{G}$ memory.
We use Intel Xeon E5-4650 CPUs.


For our benchmark test, we analyze spectrograms consisting of $151\times500$ pixels ($\unit[151]{Hz}\times\unit[250]{s}$) using the same deep-search settings used to analyze the ADI~1 waveforms in the previous section.
The computation time includes input-output tasks and other calculations, which do not take advantage of the GPU architecture.
However, these computations correspond to a tiny fraction ($\lesssim1\%$) of the total computation time.
The results are summarized in Table.~\ref{tab:gpu}.
We find that \stochsky\ calculations can be carried out $\approx10\times$ faster on GPUs than CPUs.

\begin{table}
  \begin{tabular}{|c|c|}
    \hline
    hardware & computation time \\\hline
    CPU & $\unit[3800]{s}$ \\\hline
    GPU & $\unit[380]{s}$ \\\hline
  \end{tabular}
  \caption{
    Relative computation times for \stochsky\ running on different architectures.
    The spectrogram is $151\times500$ pixels in size and we use $T=2\times10^7$ templates.
  }
  \label{tab:gpu}
\end{table}

Using our benchmark tests, we estimate the computational requirements for full-fledged gravitational searches running \stochtrack\ and \stochsky\ on GPUs.
(Interestingly, \stochtrack\ and \stochsky\ take about the same time to run given identical parameters.)
We consider two analyses: one targeted (using \stochtrack) and one all-sky (using \stochsky).
For both analyses, we assume an analysis band of $\Delta f=\unit[1200]{Hz}$ (following~\cite{lgrb}).
For the targeted analysis, we assume that search analyzes $n_\text{trig}=50$ external triggers, e.g., from gamma-ray bursts; see~\cite{lgrb}.
Following~\cite{lgrb}, we assume that the search is carried out in a $\Delta t=\unit[1500]{s}$-wide on-source window.
For the targeted analysis, we further assume that $n_\text{ts}=100$ time-shift analyses are carried out in order to evaluate the significance of candidate events; see, e.g.,~\cite{s1_cbc}.
For the all-sky analysis, we assume $n_\text{ts}=10$.

Before we present estimates of computational cost, it will be useful to define a new variable: $T_{150}$, the number of templates per $\unit[150]{Hz}$ of bandwidth.
This variable is useful since, all else equal, bigger bands must be analyzed with more templates than smaller bands due to the increased size of the template parameter space.
We chose $\unit[150]{Hz}$ in order to facilitate comparisons with the ADI~1 and ADI~2 results given in Tables~\ref{tab:mc} and~\ref{tab:rn}.
However, we note that waveforms FA~1 and FA~2 are analyzed in a $\unit[900]{Hz}$-wide band, six times wider than the ADI analysis band.
Thus, $T_{150}\approx3\times10^5$ corresponds to $T=2\times10^6$ (the default search) in the FA~1 and FA~2 analysis analysis band.
$T_{150}=2\times10^7$ corresponds to $\approx1\times10^8$ ({\em more sensitive} than \stochtrack\ $10\times$) in the FA~1 and FA~2 analysis analysis band.

Given our assumptions, the estimated computational time for a triggered \stochtrack\ search with GPUs is:
\begin{equation}\label{eq:stochtrack}
  \begin{split}
    t_c \approx & \unit[15]{days} 
    \left(\frac{T_{150}}{2\times10^7}\right) 
    \left(\frac{\Delta t}{\unit[1500]{s}}\right)
    \left(\frac{\Delta f}{\unit[1200]{Hz}}\right)
    \left(\frac{n_\text{trig}}{50}\right) \\
    & 
    \left(\frac{n_\text{ts}}{100}\right)
    \left(\frac{128}{n_\text{GPU}}\right)
  .
  \end{split}
\end{equation}
Here $n_\text{GPU}$ is the number of GPUs.
The estimated computational time for an all-sky search with GPUs is:
\begin{equation}\label{eq:stochsky}
  \begin{split}
    t_c \approx & \unit[13]{days}
    \left(\frac{T_{150}}{3\times10^5}\right)
    \left(\frac{\Delta t}{\unit[1]{yr}}\right)
    \left(\frac{\Delta f}{\unit[1200]{Hz}}\right) \\
    &
    \left(\frac{n_\text{ts}}{10}\right)
    \left(\frac{128}{n_\text{GPU}}\right)
  .
  \end{split}
\end{equation}
From Eq.~\ref{eq:stochtrack}, we conclude that GPUs can facilitate a deep-search sensitivity with \stochtrack\ using modest computational resources.
From Eq.~\ref{eq:stochsky}, we conclude that a year-long all-sky analysis with default-sensitivity \stochsky\ can also be carried out using reasonable computational resources.

Since we know that $T_{150}=3\times10^5$ \stochsky\ sensitivity can improve significantly with added templates, it would be advisable to follow up on $\approx10$ of the loudest events identified by the all-sky analysis, with a deeper $T_{150}=2\times10^7$ search.
This would add only a marginal increase to the computational burden while ensuring that a marginal detection is promoted to a strong detection (or revealed to be a noise fluctuation).

The sensitivity of an all-sky search with \stochsky\ can be increased after the analysis has commenced (supposing, for example, that more GPUs become available) through the use of intermediate data files.
Namely, we recommend recording $\text{SNR}_\text{tot}$ for each spectrogram.
If multiple runs of the analysis are carried out, one can choose the largest value of $\text{SNR}_\text{tot}$ among each run and for every spectrogram in a simple post-processing step~\footnote{Do not forget to use a different random seed for each run, good reader.}.
In other words, it is easy to combine the results from three runs with $T_{150}=3\times10^5$ in order to obtain results identical to a single $T_{150}=9\times10^5$ search.
This parallelizability can be exploited to plan for a computationally conservative analysis, while being ready for a more aggressive analysis, should the resources be available.

\section{Conclusions and Future Work}\label{conclusions}
In previous work, we proposed a new seedless clustering algorithm called \stochtrack\ and demonstrated how it could significantly improve the sensitivity of searches for long-lived, unmodeled gravitational-wave transients.
Here we extend the principle of \stochtrack\ to the case of an all-sky search, when there is no external trigger telling us where on the sky to look.
We compare the sensitivity of \stochsky\ to that of a seed-based algorithm (which takes the true sky direction as input), and find that, for the most part, \stochsky\ is significantly more sensitive, even though it is searching for the signal in a much larger parameter space.

We point out that \stochtrack\ and \stochsky\ are ``embarrassingly parallel'' algorithms and we perform benchmark tests using CPUs and GPUs.
We find that GPUs can carry out \stochtrack\ and \stochsky\ calculations ten times faster than CPUs.
We estimate the computational cost of realistic analyses, and show that interesting investigations can be carried out in a reasonable amount of time with a modest number of GPUs.

While we present \stochsky\ as a tool for all-sky analyses, it should also be very helpful in targeted analyses in which the sky localization of the external trigger is large compared to the point-spread function of the gravitational-wave detector network.
Instead of drawing the time delay variable $\Delta\tau$ (see Eq.~\ref{eq:german_sum}) from a distribution derived from an isotropic prior, it is straightforward to draw it from a distribution corresponding to a particular patch of sky.
This hybrid solution provides an efficient alternative to running \stochtrack\ for many different directions.

We previously mentioned in~\cite{stochtrack} the possibility of using \stochtrack\ to search for compact binaries.
In general, compact binaries can be well-modeled, and so it is expected that matched filtering is the optimal search strategy.
However, there are good reasons to explore alternative methods: 
\begin{itemize}
  \item Improved robustness and redundancy with an alternative method.
  \item Investigate potentially challenging corners of parameter space, e.g., systems with non-negligible spin and/or eccentricity.
  \item Detect exotic systems and/or new physics which are not included in matched filter template banks.
\end{itemize}

\begin{figure*}[hbtp!]
  \begin{tabular}{cc}
    \psfig{file=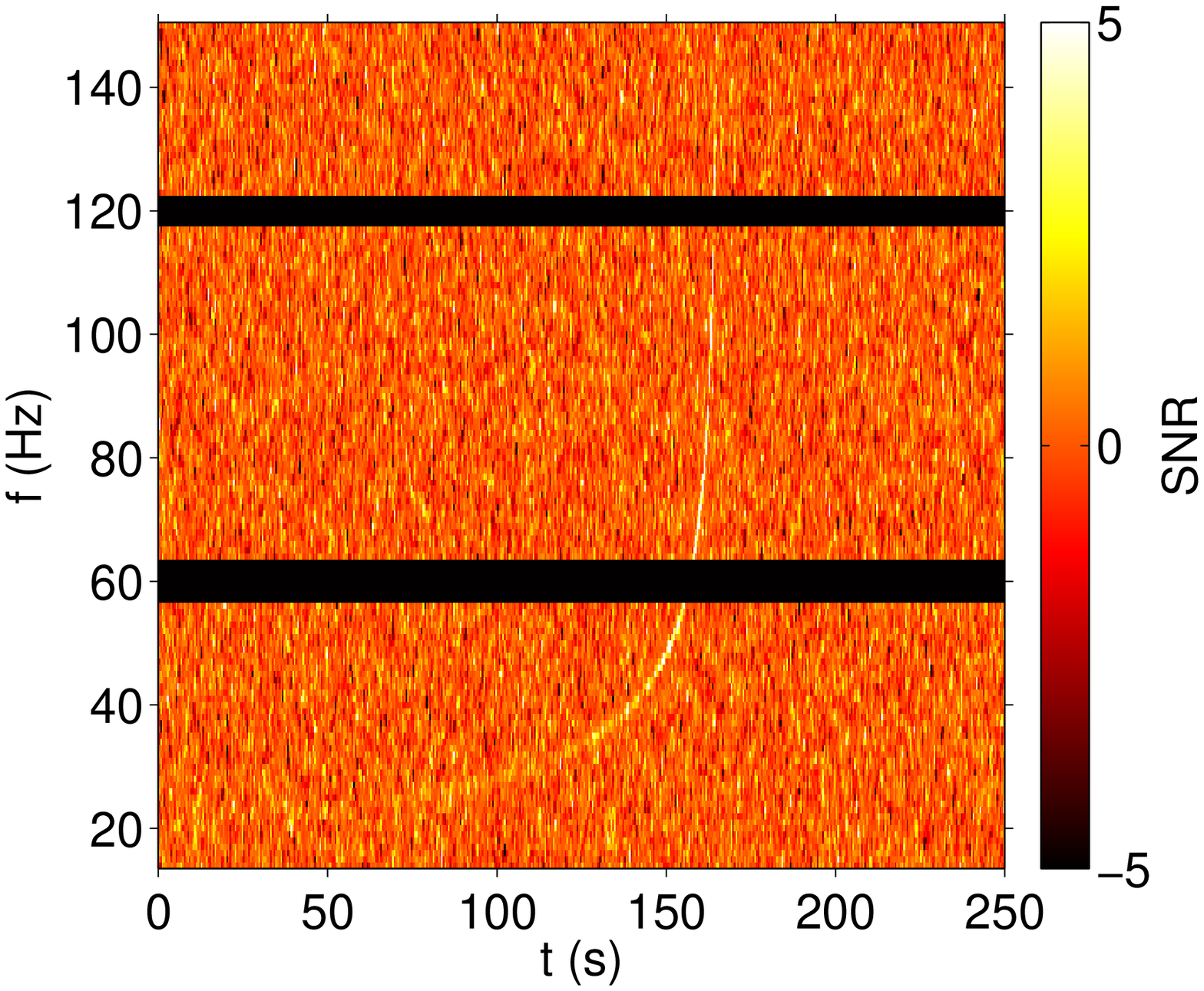, height=2.8in} &
    \psfig{file=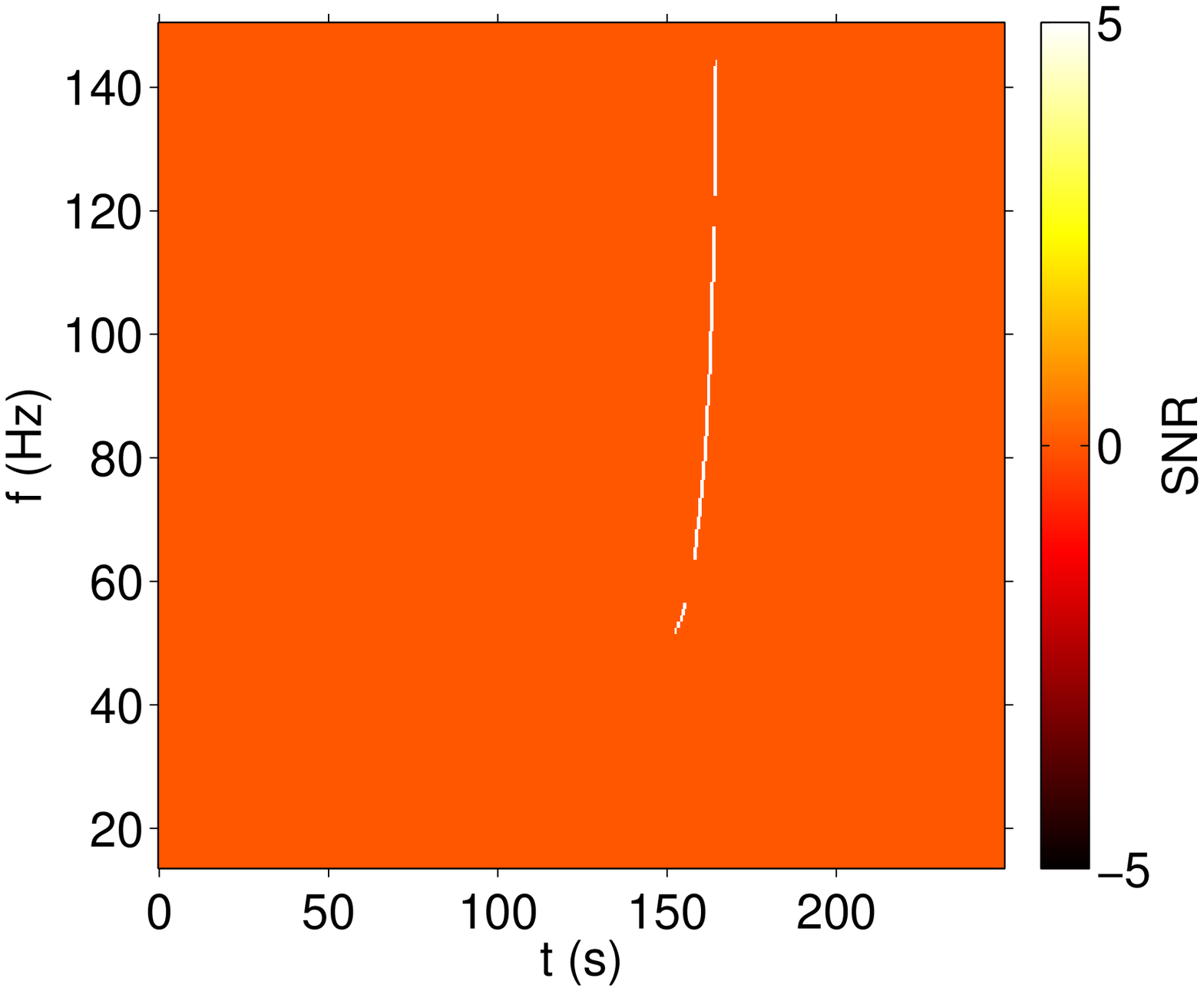, height=2.8in}
  \end{tabular}
  \caption{
    Left: $\text{SNR}$ spectrogram for a (very loud) binary neutron star signal in Monte Carlo noise.
    Right: the signal recovered by \stochtrack.
    The \stochtrack\ algorithm can detect binary neutron star signals in Advanced LIGO Monte Carlo noise with with $\text{FAP}=0.1\%$ and $\text{FDP}=50\%$ at a distance of $\unit[160]{Mpc}$.
    Note that \stochsky\ does not do a good job of catching the beginning of the signal.
    This is not due to a lack of templates, but rather because the binary neutron star signal is not especially well-described with a second-order B\'ezier curve.
  }
  \label{fig:bns}
\end{figure*}

In order to place this discussion in context and to motivate future work, we close by reporting the results of a sensitivity study for detecting the coalescence of two $1.4 M_\odot$ neutron stars with \stochtrack.
We consider the case of an optimally oriented system at an optimal sky location.
We assume the signal is confined to a $\unit[660]{s}$ on-source region as in previous searches triggered by gamma-ray bursts~\cite{s6grb}.
We find that such a binary neutron star coalescence can be detected in Advanced LIGO Monte Carlo noise using \stochtrack\ with $\text{FAP}=0.1\%$ and $\text{FDP}=50\%$ at a distance of $\unit[160]{Mpc}$.
For comparison, the best $90\%$ upper limit from initial LIGO and Virgo on binary neutron star coalescence coincident with gamma ray bursts is $\unit[37]{Mpc}$~\cite{s6grb}.
It is probable that the sensitivity of \stochtrack\ to binary neutron stars can be enhanced with additional tuning; see Fig.~\ref{fig:bns}.
Thus, the application of \stochtrack\ and \stochsky\ to compact binary coalescence signals appears promising and worthy of future work.

\begin{acknowledgments}
We thank Stuart Anderson, Juan Barayoga, and Fred Donovan for assistance with GPUs.
We thank Anthony Piro for sharing the fallback accretion waveforms used in this analysis.
We thank Tanner Prestegard for helpful comments on a draft of this paper.
ET is a member of the LIGO Laboratory, supported by funding from United States National Science Foundation.
LIGO was constructed by the California Institute of Technology and Massachusetts Institute of Technology with funding from the National Science Foundation and operates under cooperative agreement PHY-0757058.
MC is supported by National Science Foundation Graduate Research Fellowship Program, under NSF grant number DGE 1144152.
This paper has been assigned document number LIGO-P1400010.
\end{acknowledgments}

\begin{appendix}
\section{Formalism}\label{formalism}
The signal-to-noise ratio spectrogram can be written as
\begin{equation}
  \rho(t;f|\hat\Omega) = \widehat{Y}(t;f|\hat\Omega) / \widehat\sigma(t;f|\hat\Omega) .
\end{equation}
Here $\widehat{Y}$ is an estimator for cross-power:
\begin{equation}
  \widehat{Y}(t;f|\hat\Omega) =  \frac{2}{\cal N} \text{Re} \left[
      Q_{IJ}(t;f|\hat\Omega) \, \tilde{s}^*_I(t;f)  \tilde{s}_J(t;f)
      \right] ,
\end{equation}
and $\widehat\sigma^2$ is an estimator for its variance
\begin{equation}
  \widehat\sigma^2(t;f|\hat\Omega) =  \frac{1}{2} \left| Q_{IJ}(t;f|\hat\Omega) \right|^2
    P'_I(t;f) P'_J(t;f) .
\end{equation}
Here ${\cal N}$ is the normalization from a discrete Fourier transform and $Q_{IJ}(t;f|\hat\Omega)$ is a filter function, which accounts for the time delay between detectors $I$ and $J$ as well as the detector responses.
Typically $Q_{IJ}(t;f|\hat\Omega)$ is defined such that $\hat{Y}_{IJ}(t;f)$ is an unbiased estimator for gravitational-wave power~\cite{stamp}.
The variables $P'_I(t;f)$ and $P'_J(t;f)$ are the auto-power spectral densities for detectors $I$ and $J$ in the segments neighboring $t$.

It follows that
\begin{equation}
  \lambda(t;f) = \frac{1}{\cal N} \sqrt{\frac{2}{P'_I(t;f) P'_J(t;f)}}
\end{equation}
For additional details, the reader is referred to~\cite{stamp}.

\section{Model Parameters}\label{parameters}
This section reproduces details about the test waveforms from~\cite{stochtrack}.
The FA waveforms~\cite{piro:11,pirothrane12} are described by the following parameters: initial protoneutron star mass $M_0$, maximum neutron star mass $M_\text{max}$, a dimensionless factor related to the supernovae explosion energy $\eta\approx0.1$--$10$, and the radius of the protoneutron star $R_0$.
The values of these parameters for FA~1 and FA~2 are given in Table~\ref{tab:FA}.
The ADI waveforms~\cite{lucia} are parameterized by black hole mass $M_\text{BH}$, dimensionless spin parameter $\alpha^\star=[0,1)$, the fraction of the accretion disk mask that forms clumps $\epsilon\approx0.01$--$0.2$, and the torus mass $m$.
The values of these parameters for ADI~1 and ADI~2 are given in Table~\ref{tab:ADI}.

\begin{table}
  \begin{tabular}{|c|c|c|c|c|}
    \hline
    waveform & duration (s) & $f_\text{min}$--$f_\text{max}$ (Hz) &
    $\delta{t}\times\delta{f}$ & $t_\text{min}$ \\\hline
    ADI 1 & $39$ & $130$--$170$ &
    $\unit[1]{s}\times\unit[1]{Hz}$ & $\unit[35]{s}$ \\\hline
    ADI 2 & $230$ & $110$--$260$ &
    $\unit[1]{s}\times\unit[1]{Hz}$ & $\unit[100]{s}$ \\\hline
    FA 1 & $25$ & $1170$--$1530$ &
    $\unit[0.5]{s}\times\unit[2]{Hz}$ & $\unit[20]{s}$ \\\hline
    FA 2 & $200$ & $790$--$1080$ &
    $\unit[1]{s}\times\unit[1]{Hz}$ & $\unit[100]{s}$ \\\hline
  \end{tabular}
  \caption{
    A summary of the waveforms used in our sensitivity study from~\cite{stochtrack}.
    The second and third columns describe the duration and frequency range of the waveform respectively.
    The fourth column gives the spectrogram resolution used to analyze each waveform.
    The fifth column specifies the minimum signal duration assumed in each search.
    The ADI waveforms are down-chirping accretion-disk instability waveforms~\cite{vanputten:01,vanputten:08,lucia} whereas the FA waveforms are up-chirping fallback accretion powered waveforms~\cite{pirothrane12,piro:11}.
  }
  \label{tab:waveforms}
\end{table}

\begin{table}[h]
  \begin{tabular}{|c|c|c|c|c|}
    \hline
    waveform & $M_0$ ($M_\odot$) & $M_\text{max}$ ($M_\odot$) & $\eta$ & $R_0$ (km) \\\hline
    FA 1 & $1.3$ & $2.5$ & $10$ & $20$ \\\hline
    FA 2 & $1.3$ & $2.5$ & $1$ & $25$ \\\hline
  \end{tabular}
  \caption{Parameters for FA waveforms from~\cite{stochtrack}.  See~\cite{pirothrane12} for additional details.}
  \label{tab:FA}
\end{table}

\begin{table}[h]
  \begin{tabular}{|c|c|c|c|c|}
    \hline
    waveform & $M_\text{BH}$ ($M_\odot$) & $\alpha$ & $\epsilon$ & $m$ ($M_\odot$) \\\hline
    ADI 1 & $5$ & $0.3$ & $0.05$ & $1.5$ \\\hline
    ADI 2 & $10$ & $0.95$ & $0.04$ & $1.5$ \\\hline
  \end{tabular}
  \caption{Parameters for ADI waveforms from~\cite{stochtrack}.  See~\cite{lucia} for additional details.}
  \label{tab:ADI}
\end{table}

\end{appendix}

\bibliography{stochsky}

\end{document}